\documentclass[pra,aps,showpacs,singlecolumn,amsmath,amssymb,floatfix]{revtex4}
\usepackage{graphicx}
\usepackage{amssymb,amsmath}
\usepackage{subfigure}
\usepackage{bbm}
\usepackage[usenames]{color}
\begin{document}

\title{Grey solitons in a strongly interacting superfluid Fermi Gas}
\author{Andrea Spuntarelli$^{1}$, Lincoln D. Carr$^{2}$, Pierbiagio Pieri$^{1}$, and Giancarlo C. Strinati$^{1}$}
\affiliation{$^1$Dipartimento di Fisica, Universit\`{a} di Camerino, I-62032 Camerino, Italy\\
$^2$Department of Physics, Colorado School of Mines, Golden, Colorado 80401, U.S.A.}
\date{\today}

\begin{abstract}
The Bardeen-Cooper-Schrieffer to Bose-Einstein condensate (BCS to BEC) crossover problem is solved for stationary grey solitons via the Boguliubov-de Gennes equations at zero temperature. These \emph{crossover solitons} exhibit a localized notch in the gap and a characteristic phase difference across the notch for all interaction strengths, from BEC to BCS regimes.  However, they do not follow the well-known Josephson-like sinusoidal relationship between velocity and phase difference except in the far BEC limit: at unitarity the velocity has a nearly linear dependence on phase difference over an extended range.  For fixed phase difference the soliton is of nearly constant depth from the BEC limit to unitarity and then grows progressively shallower into the BCS limit, and on the BCS side Friedel oscillations are apparent in both gap amplitude and phase.  The crossover soliton appears fundamentally in the gap; we show, however, that the density closely follows the gap, and the soliton is therefore observable.  We develop an approximate power law relationship to express this fact: the density of grey crossover solitons varies as the square of the gap amplitude in the BEC limit and a power of about 1.5 at unitarity.
\end{abstract}


\maketitle

\section{Introduction}
\label{sec:introduction}

Ultracold quantum gases present a highly tunable system in which to explore the physics of superconductivity and superfluidity.  In fact, by tuning the interactions of an ultracold Fermi gas via a Fano-Feshbach resonance, one can crossover continuously between Bardeen-Cooper-Schrieffer (BCS) $s$-wave superconductivity, in which the fermions form large overlapping Cooper pairs, and a Bose-Einstein condensate (BEC) of tightly bound Fermi pairs.  In between these two regimes, when the pairing length is on the order of the atomic spacing, one obtains a unitary gas, an exciting state of quantum matter under intensive exploration~\cite{kinast2004,regal2004,bartenstein2004b,chenQ2005b}.  The unitary gas is an example of a strongly correlated quantum fluid, and has close analogies in quark-gluon plasmas~\cite{jacak2010}, cold neutral plasmas~\cite{killian2010}, graphene~\cite{mueller2009}, and anti-de Sitter theory via holographic dualities~\cite{johnsonCV2010}.  In ultracold quantum gases, a smoking gun signal of Cooper pairing is quantized rotation localized in vortices (similar to quantized magnetic field in a superconducting metal).
For a superfluid Fermi gas, such vortices are associated primarily with the superfluid gap parameter $\Delta$, but have been directly observed through their effects on the density profiles for the whole crossover from BEC to unitary gas to BCS~\cite{zwierlein2005a}, and are an emergent property of this many-body quantum system.

In one dimension (1D) the key emergent property and analogous structure to a vortex is a \emph{dark soliton}.  A dark soliton is a persistent robust localized nonlinear wave that appears as a density notch in BECs. Unlike their higher dimensional vortex analogs, dark solitons can continuously transform from a deep notch with a node, called a \emph{black} soliton, to a \emph{grey} soliton with a progressively shallower notch, all the way to a constant density. Dark solitons exist in higher dimensions as well, where they take the form of dark bands in 2D and dark planes in 3D~\cite{carr2008f,carr2008g}.  Dark solitons in 1D and 3D have been observed in BEC experiments on atomic bosons~\cite{burger1999,denschlag2000,anderson2001,ginsberg2005,weller2008,stellmerS2008} but so far remain unobserved in atomic Fermi systems. In a superfluid Fermi gas, solitons, like vortices, appear fundamentally in the superfluid gap parameter, but may produce observable effects also on the density. We term dark solitons in Fermi gases \emph{crossover solitons}.  In the BEC case, mean field theory has proved to closely match observations~\cite{denschlag2000,feder2000}; modified mean field theories treating thermal~\cite{jackson2006,jacksonB2007,martinAD2010} and/or quantum fluctuations~\cite{dziarmaga2003,martinAD2010,gangardt2010} as well as entangled quantum many-body numerical methods~\cite{carr2009l,carr2009m} have predicted dark soliton quantum delocalization and both thermal and quantum decay.  The dark soliton excitation spectrum in a unitary gas has been studied via holographic duality~\cite{keranen2009,keranen2009b}.  However, for treating solitons and vortices through the whole BCS-BEC crossover the only numerically tractable method to date is solution of the Boguliubov-de Gennes equations (BdGE), as three of us have presented in a complete review~\cite{spuntarelli2010}; soliton solutions of modified BdGE also appear in the context of the chiral Gross-Neveu model used to treat chiral superconductors~\cite{basar2008a,basar2008b}.

Recently, we used the methods of Ref.~\cite{spuntarelli2010} to study the stationary Josephson effect through the BCS-BEC crossover~\cite{spuntarelli2007}. The evolution through the crossover of the Josepshon characteristic, {\em i.e.}, the relationship between the Josephson current and the phase difference across a given barrier, was determined in this way. The Josephson current was found to be enhanced at about unitarity for all barriers, and the mechanism behind such an enhancement was explained in terms of the behavior of the Landau critical velocity through the crossover.  Our colleagues at Trento used related methods~\cite{antezza2007} to treat the crossover soliton for the special case of a black, or static soliton, sometimes called a kink, at which the gap parameter has a node at its minimum; in this case the solution of the BdGE involves only real wave functions.  In the more general case of \emph{grey} (moving) solitons, the soliton velocity has a special relation to the phase difference over the notch: the phase decreases monotonically from $\pi$ to 0 as the velocity increases from zero to the local speed of sound $c_s$, and the BEC mean field theory, called the nonlinear Schr\"odinger equation (NLSE) or Gross-Pitaevskii equation, has an exact solution that describes both density and phase via Jacobi elliptic functions~\cite{carr2000a,kivshar1998}.  All these intriguing features of grey solitons remain unexplored.

In this Article we treat the general case of a crossover soliton, from BCS to BEC regimes, using our numerical methods, which were well established in the context of the Josephon effect problem~\cite{spuntarelli2007,spuntarelli2010}, to find complex single soliton solutions.  We find that crossover solitons display a localized notch in the gap amplitude, and a characteristic phase difference in the gap over all interaction regimes, with Friedel oscillations apparent in both amplitude and phase on the BCS or fermionic side.   Although it is the gap that describes the crossover soliton amplitude and phase in analogy to the dark soliton solution of the NLSE, experiments in fact observe the density.  Therefore we provide an analysis of the difference between gap and density and demonstrate that crossover solitons are also evident in the density. In particular, we find that for a unitary Fermi gas grey solitons may present even better contrast in the density than in the gap profile, at variance with what found for black solitons~\cite{antezza2007}.
We observe that crossover solitons deviate strongly from the well-known Josephson-like relation between velocity and phase difference which holds for BEC solitons, and that these deviations are most pronounced at unitarity. In addition, for fixed phase difference $\delta\phi$ the notch is of nearly constant depth from the BEC limit to unitarity and then grows shallower as the BCS limit is approached, in both gap and density.

We emphasize that we treat infinite 3D fermionic systems.  What is 1D in these studies is the profile of the solitary wave, which is translationally invariant in the two remaining spatial directions, as discussed in detail in the presentation of our methods in Sec.~\ref{sec:method}.  This choice was made to be close to the current experiments on the BCS-BEC crossover with ultracold Fermi gases, which have been performed almost invariably for 3D systems.
In addition, in this first exploration of crossover solitons we wanted to eliminate, as much as possible, finite-size effects from the calculations.  In Sec.~\ref{sec:results} we show the results of our numerical calculations.  Finally in Sec.~\ref{sec:conclusions} we discuss our results and conclude.

\section{Method}
\label{sec:method}

We study the behavior at zero temperature of a moving soliton in a system of neutral fermions, mutually interacting via an attractive short-range potential parameterized in terms of the scattering length $a_{F}$. The dimensionless strength of the interaction is thus given by the coupling parameter $(k_{F} a_{F})^{-1}$ where $k_{F}$ is the Fermi wave vector.
This parameter drives the BCS-BEC crossover and ranges from $(k_{F} a_{F})^{-1} \ll -1$ in the weak-coupling (BCS) limit to
$1 \ll (k_{F} a_{F})^{-1}$ in the strong-coupling (BEC) limit, while the interaction range
$-1 \lesssim (k_{F} a_{F})^{-1} \lesssim +1$ is termed the crossover region.  The special case $(k_{F} a_{F})^{-1} = 0$ is known as the unitary limit or unitary gas, and is the subject of extensive study via a variety of analytical and numerical methods, as mentioned in Sec.~\ref{sec:introduction}.

We consider specifically a homogeneous system, extending to infinity in the three spatial dimensions. We look for
solitonic solutions of the gap profile $\Delta(\mathbf{r},t)$ which are translationally invariant along the $y$ and $z$
directions. The gap profile for a soliton moving with velocity $v$ along the $x$ axis will be thus given by \begin{equation}
\Delta(\mathbf{r},t)=\Delta(x - v t).\label{eq:first}
\end{equation}
As we will treat only a single soliton, formally our solutions should be called solitary waves until elastic soliton-soliton interactions are demonstrated throughout the crossover.

Equation~(\ref{eq:first}) describes a soliton propagating with a velocity $v$ in a superfluid at rest.  In order to work with a stationary problem, it is convenient to make a
Galilean transform to the frame of reference co-moving with the soliton.
In such a frame the soliton profile is time independent and the superfluid is flowing with constant velocity $-v$.
We define $v>0$ for a soliton moving from right to left, such that in the co-moving frame the superfluid
is flowing from left to right, consistently with our previous work on the Josephson effect.
The gap profile will thus transform to
\begin{equation}
\Delta(\mathbf{r})=\Delta(x) e^{2 i q x},
\end{equation}
with $q=-m v$, $m$ being the fermion mass, as dictated by
Galilean invariance.
Besides the complex oscillating phase $2 q x$ resulting from the Galilean boost, $\Delta(\mathbf{r})$
has an additional phase dependence $\phi(x)$ which is associated with the finite velocity
of the soliton in the original frame where the soliton is moving. We thus set
\begin{equation}
\Delta(x)= |\Delta(x)| e^{i \phi(x)}.
\end{equation}
The counterflow between background current and the dark soliton is a direct extension from the NLSE, as described in Ref.~\cite{carr2000a}.

For a static or black soliton the phase $\phi(x)$ jumps by a factor $\pi$ at the soliton notch, where $\Delta(x)$
vanishes. For a moving soliton, the notch does not have a node, and the phase $\phi(x)$ changes instead continuously between the two different bulk values $\phi(x=+\infty)$ and $\phi(x=-\infty)$ reached
away from the notch; the soliton is called grey to describe its reduced density contrast.
Quite generally, there is a correspondence between the overall phase difference
\begin{equation}
\delta \phi=\phi(x=+\infty)-\phi(x=-\infty)
\end{equation}
and the soliton velocity $v$.
In the BEC limit, where the fermionic BdGE reduce to the NLSE for composite bosons~\cite{pieri2003} with mass $m_B= 2 m$,
scattering length $a_B= 2 a_F$ and condensate amplitude
\begin{equation}
\Phi(x) = \Delta(x) \sqrt{\frac{m^{2} a_{F}}{8 \pi}}\,,
\end{equation}
one finds the Josephson-like relation
\begin{equation}
v = v_c \cos(\delta\phi/2),
\label{eq:josephson}
\end{equation}
where $v_c$ is the sound velocity~\cite{reinhardt1997}.  We clarify that the ``standard'' Josephson effect occurs in a weak tunneling regime which is achieved for sufficiently high barriers. In this regime one replaces the $\cos(\delta\phi/2)$ function in Eq.~(\ref{eq:josephson}) with a $\sin(\delta\phi)$ function.  As the barrier is lowered the $\sin(\delta\phi)$ dependence gradually turns into a $\cos(\delta\phi/2)$ dependence.  In using the term ``Josephson-like'' we allude to our present case being that of a vanishing barrier.
In Eq.~(\ref{eq:josephson}) the soliton velocity ranges from $0$ with the associated phase difference value $\delta \phi=\pi$ for the static case, to the sound
velocity $v_c$ and the corresponding phase difference value $\delta\phi=0$ in the limit where the grey soliton is actually an infinitesimal perturbation propagating
with velocity $v_c$.
More generally, in the BCS-BEC crossover
\begin{equation}
v=v_c((k_F a_F)^{-1}) f(\delta \phi),
\end{equation}
with $f(0)=1$ and $f(\pi)=0$.  In the former limit, $f(0)=1$, and $v=v_c((k_F a_F)^{-1})$  is the Landau critical velocity
for the breakdown of superfluidity in a homogeneous superfluid, determined by the pair-breaking or by the sound-mode velocity on the two
sides of the crossover, as we have previously described~\cite{spuntarelli2007}.  The limiting tendency of our numerical calculations in Sec.~\ref{sec:results} supports this interpretation.  In the latter limit, $f(\pi)=0$, the black soliton solution is reproduced, as previously studied in~\cite{antezza2007}.  Thus both the infinitesimal notch and the notch with a node have a clear interpretation in terms of previous results.

In practice, the order parameter $\Delta(\mathbf{r})$ is obtained by solving the BdGE at zero temperature for the two-component fermionic wave functions~\cite{deGennes1989},
\begin{equation}
\left(
\begin{array}{cc}
\mathcal{H}(\mathbf{r}) & \Delta(\mathbf{r})            \\
\Delta(\mathbf{r})^{*}  & - \mathcal{H}(\mathbf{r})
\end{array}
\right)
\left( \begin{array}{c}
u_{\nu}(\mathbf{r}) \\
v_{\nu}(\mathbf{r})
\end{array}
\right)
= \epsilon_{\nu}
\left( \begin{array}{c}
u_{\nu}(\mathbf{r}) \\
v_{\nu}(\mathbf{r})
\end{array}
\right) ,                                    \label{B-dG-equations}
\end{equation}
with
\begin{equation}
\mathcal{H}(\mathbf{r}) = -  \frac{\nabla^{2}}{2m} + V(\mathbf{r}) - \mu ,
\end{equation}
where $m$ and $\mu$ are the fermion mass and chemical potential.
The function $\Delta(\mathbf{r})$ is determined via the \emph{self-consistency condition},
\begin{equation}
\Delta(\mathbf{r}) = g \sum_{\nu} u_{\nu}(\mathbf{r})
v_{\nu}(\mathbf{r})^{*},       \label{self-consistency}
\end{equation}
where $- g$ is the ``bare'' strength of the local fermionic attraction, which is eliminated eventually in favor of the scattering length
$a_{F}$ with an appropriate regularization procedure~\cite{spuntarelli2010}.
The particle number density $n(\mathbf{r})$ is in turn determined by the equation,
\begin{equation}
n(\mathbf{r}) = 2 \sum_{\nu}
|v_{\nu}(\mathbf{r})|^2.       \label{density}
\end{equation}

We determined the self-consistent solution of the BdGE via the same numerical method adopted in our previous work~\cite{spuntarelli2007}, which in turn extended the method
introduced in Ref.~\cite{riedel1996} for the study of the Josephson effect in a weak coupling superconductor to the study of the same problem in a strongly-interacting
superfluid Fermi gas.
We briefly recall here the method adopted to solve this problem, and refer the reader to our review~\cite{spuntarelli2010} for a complete and thorough description.
The profile of $\Delta(x)$ is made piecewise constant over a dense number of intervals, typically $100$. In this way, in each interval the eigenfunctions
$u_{\nu}(\mathbf{r}), v_{\nu}(\mathbf{r})$ of the BdGE are plane waves. The wave functions in contiguous intervals are
connected by continuity conditions.
In general, a complete and orthonormal set of solutions of the BdGE is made by both discrete (bound) states, known as Andreev-Saint James
states \cite{andreev1964,saint-james1964}, and continuum states.  For the continuum states we choose
outgoing boundary conditions for waves impinging on the barrier from the left and right in order to single out a complete and orthonormal set of eigenfunctions.
Self-consistency for the gap profile is implemented from the outcomes of such
a scattering problem over a less dense grid of points, typically $20$.
The translational invariance associated with the position of the soliton notch is removed in the numerical calculation by enforcing the condition $\Delta(-x)=\Delta(x)^*$, up to a global phase factor eliminated by setting
$\phi(-\infty)=0$ (in the context of the Josephson effect the translational symmetry is instead explicitly broken by the presence of a barrier).

The solitonic solutions of the BdGE are obtained by fixing the phase difference $\delta \phi$ at a given value in the
range $[0,\pi]$ and determining the associated soliton profile and velocity $v$ through the self-consistent solution of the BdGE.
In the extreme BEC limit $(k_F a_F)^{-1}\to \infty$ one can prove that the BdGE become the NLSE~\cite{pieri2003}, and for the NLSE all soliton properties are well-established~\cite{kivshar1998,carr2000a}.  In the BCS limit $(k_F a_F)^{-1}\to -\infty$ our numerical method develops large error bars, and in practice we are limited to the near-unitary region.  This is due to the difficulty in dealing simultaneously with two
length scales: $k_F^{-1}$, determining Friedel oscillations; and the healing length $\xi$, which diverges in the weak-coupling limit.  For the Josephson problem, i.e., in the presence of a finite barrier,
the problems in weak-coupling are milder~\cite{spuntarelli2007,spuntarelli2010}.  It is known that for a sufficiently high barrier self-consistency becomes less and less important in the weak-coupling limit in determining the Josephson characteristic, i.e., the
Josephson current $j$ vs.~the phase difference $\delta\phi$, through the crossover.  For finite barriers, this helped
when solving the equations.  In our present study, our numerical methods are limited to the interaction regimes $-0.5 \lesssim (k_F a_F)^{-1} \lesssim 1$.   Specifically, our error is determined by varying the number of grid points used in the solution of the scattering problem, the smaller number of grid points used to enforce self-consistency, and the overall box size.  In general all calculations presented in Sec.~\ref{sec:results} are accurate to a few percent, and therefore on the order of the point size in plots (in Fig.~\ref{fig:5}(c) we reduce the point size and explicitly show the actual error bars).

Finally, we briefly clarify how our present study relates to previous work.  In Ref.~\cite{spuntarelli2007} we studied the Josephson characteristic. We  thus studied the
flow of a superfluid across a potential barrier, and determined for
given barriers the curves $j(\delta\phi)$ through the crossover.
In the present case the barrier is absent. However, in the reference frame
co-moving  with the soliton, the solitonic solution studied here can
be seen as the limit for a vanishing barrier of the Josephson problem studied
in our previous work.  On the other hand, in Ref.~\cite{antezza2007} the problem of a static soliton in a superfluid Fermi gas was studied for purely real solutions (up to a trivial phase) with a node. Our work extends the analysis of  Ref.~\cite{antezza2007} by allowing the soliton to move. This requires
one to consider {\em complex} solutions of the gap parameter.  The soliton velocity determines the soliton profile, the depth of the soliton notch in both gap and density, and the phase difference, as we describe in the following section.

\section{Results}
\label{sec:results}

\begin{figure}[t]
\centering
\includegraphics[width=0.6\textwidth]{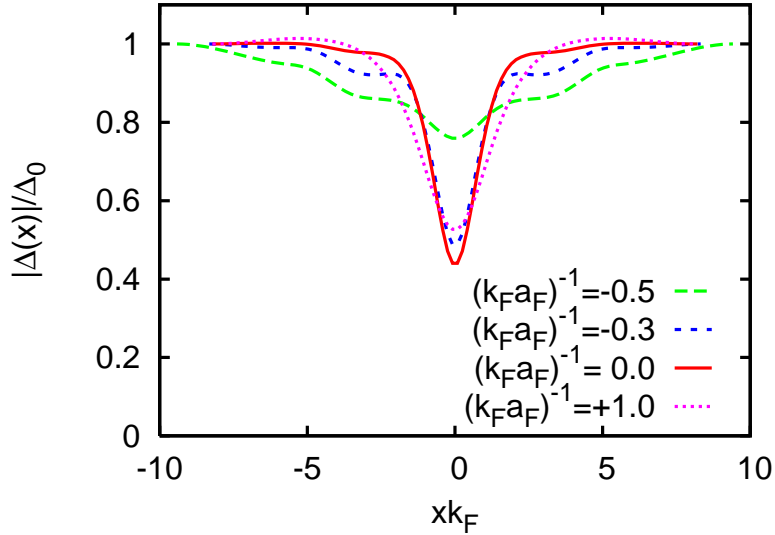}
\caption{Crossover soliton gap amplitude $|\Delta(x)|$ normalized to its bulk value  $\Delta_0$ for different coupling values at
a fixed phase difference $\delta\phi =0.6 \pi$.\label{fig:1}}
\end{figure}
Following the methods of Sec.~\ref{sec:method}, we illustrate the basic properties of crossover solitons. Figure~\ref{fig:1} shows the  soliton gap amplitude  $|\Delta(x)|$, normalized to its bulk value $\Delta_0$,  at an
intermediate value of the phase difference, $\delta\phi =0.6 \pi$, for which
$v\approx v_c/2$.  The soliton is narrowest and deepest at
unitarity, reflecting the fact that the coherence length reaches its minimum in the intermediate coupling region (see~\cite{pistolesi1996}). Friedel oscillations become clearly visible when approaching the weak-coupling region, and the soliton grows increasingly shallow as $(k_F a_F)^{-1}\to -\infty$.  One way to understand why the soliton should grow shallow is to make the conjecture that width and depth are tied together; we know this is the case in the BEC limit~\cite{kivshar1998} and we generally find that crossover soliton properties vary continuously from BEC to BCS limits, albeit with exponents which are a function of $(k_F a_F)^{-1}$ (see Table~\ref{table} and Fig.~\ref{fig:3} below for an exploration of such exponents).  We also know that $v_c\propto \Delta$ in the BCS limit and the Landau critical velocity thus drops off exponentially as $(k_F a_F)^{-1}\to -\infty$, where $v_c$ is the Landau or depairing velocity~\cite{spuntarelli2007}.  The gap $\Delta$ controls the coherence length and this in turn controls the soliton width.  Thus we can argue that as the crossover soliton traverses unitarity and moves into the BCS regime the width should get very large and the depth very shallow, even for fixed phase; moreover, this behavior can be seen as controlled by the Landau velocity.  As evident in Fig.~\ref{fig:1}, the crossover soliton also becomes shallower and broader as $(k_F a_F)^{-1}\to \infty$, although the effect is less pronounced, since the Landau velocity in the strongly interacting or BEC regime is the sound velocity, which falls off less rapidly than the depairing velocity.

\begin{figure}[t]
\centering
\includegraphics[width=0.6\textwidth]{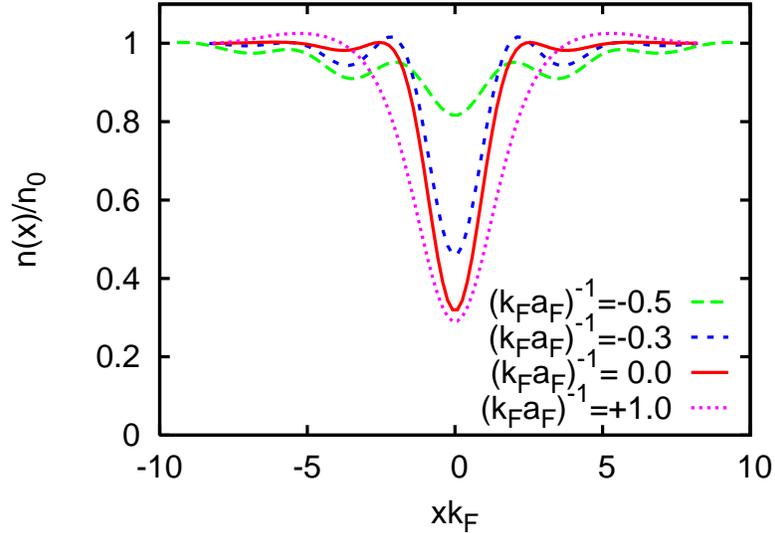}
\caption{Crossover soliton density $n(x)$ corresponding to the soliton gap amplitude shown in Fig.~\ref{fig:1}. The density is normalized to the bulk value $n_0$.\label{fig:2}}
\end{figure}
\begin{table}[]
\caption{Comparison of the depth of the soliton notch as it appears in the gap vs. the density at
fixed phase difference $\delta\phi =0.6 \pi$.} 
\centering
\begin{tabular}{c |c |c |c}
\hline\hline
Coupling $\:$ & $\:$ Gap Notch Depth $\:$ & $\:$ Density Notch Depth$\:$ & Exponent $\alpha$ in   \\
$(k_F a_F)^{-1}$ & $1-|\Delta(x=0)/\Delta_0|$ & $1-n(x=0)/n_0$  & $n(x=0)/n_0 = |\Delta(x=0)/\Delta_0|^{\alpha}$  \\ [0.5ex]
\hline
-0.5     & 0.25  & 0.28  & 0.74\\
-0.3     & 0.47  & 0.54  & 1.07 \\
0        & 0.54  & 0.68  & 1.38 \\
0.5      & 0.51  & 0.72  & 1.79 \\
1.0      & 0.47  & 0.71  & 1.93 \\
$\infty$ & 0.413 & 0.655 & 2.00 \\ [1ex] 
\hline
\end{tabular}
\label{table}
\end{table}
Experiments measure the density, not the gap.  Figure~\ref{fig:2} shows the density profile $n(x)$ of crossover solitons. In Table~\ref{table} we list the depths of the crossover soliton notch as reflected in both gap and density.  These numerical values are accurate to a few percent; our method of determining accuracy is given in Sec.~\ref{sec:method}.  As the weak-coupling BCS limit is approached one expects the spatial variations of $\Delta(x)$ to progressively affect the density profile less, as can also be seen from
Fig~\ref{fig:2}.
Still, for all the coupling values considered in the present work, the solitonic solution for the gap profile produces a sizable effect also on the density profile, thus making the soliton visible also in this quantity of direct access to experimentalists.  Note also that the Friedel oscillations in the density profile are even more evident than in the gap profile.  For the limiting value of $(k_F a_F)^{-1} \to \infty$ in Table~\ref{table} we use the analytic expressions for the BEC soliton, $\Delta(x=0)/\Delta_0=v/v_c$, together with Eq.~(\ref{eq:josephson}) and our fixed value of $\delta\phi=0.6$ chosen as a representative grey soliton example.

\begin{figure}[t]
\centering
\includegraphics[width=0.6\textwidth]{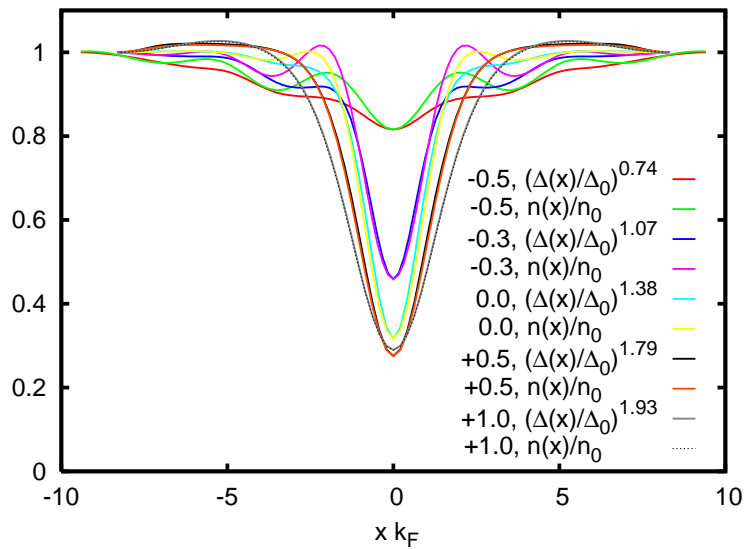}
\caption{A comparison of gap and density to determine their power law relationship as a function of $(k_F a_F)^{-1}$.  We choose the power of $\Delta(x)/\Delta_0$ which best matches the density $n(x)/n_0$ at the minimum.  In the key is listed the value of $(k_F a_F)^{-1}$ on the left and the curve being plotted on the right.  We hold $\delta\phi=0.6$ to match Figs.~\ref{fig:1} and~\ref{fig:2}.  \label{fig:3}}
\end{figure}
The simple relation of proportionality between $n(x)$ and $|\Delta(x)|^2$ which is valid in the BEC limit does not hold for the crossover soliton, as can be seen roughly from Figs.~\ref{fig:1} and Fig.~\ref{fig:2}. An attempt to extend this relationship away from the strict BEC limit for a typical grey soliton is made in Table~\ref{table}, where we determine the exponent $\alpha$ in
\begin{equation}
\frac{n(x)}{n_0} =\left|\frac{\Delta(x)}{\Delta_0}\right|^{\alpha}
\label{eq:alpha}
\end{equation}
by matching the depth of the minimum occurring in the two curves at $x=0$.  As can be seen from  Fig.~\ref{fig:3} this approximate relationship is reasonably good in an extended coupling range for grey solitons, and becomes progressively more accurate as the system is tuned to the BEC side of the resonance.
The fact that the exponent $\alpha$ approaches the expected value of 2 when moving towards the BEC limit is also a good sign for the accuracy of our algorithm.  We find that $\alpha$ changes by more than a factor of two over the range we consider, and is therefore highly variable depending on interaction strength.  Exactly on resonance we find a value for the exponent $\alpha$ close to 1.5 ($\alpha=1.38$).
Although in principle we don't expect the
local density approximation (LDA) to necessarily be correct for the soliton problem, if LDA was valid the expected exponent would indeed be 1.5, since
$\Delta(x)/E_F(x)$ should be constant within LDA exactly at unitarity, and
$E_F(x)\propto [n(x)]^{2/3}$, where $E_F(x)$ is the local Fermi energy in LDA.  We point out that for a black
soliton the gap is bound to vanish at $x=0$ for all couplings,
while the density vanishes at $x=0$ only in the extreme BEC limit~\cite{antezza2007}.  Thus for a black soliton Eq.~(\ref{eq:alpha}) cannot work in a region close to the gap notch.   In this case one should either
try another relationship or make a fit and exclude a small region close to the origin.  However, for grey solitons the relationship is useful for experiments trying to relate the observable quantity, namely density, back to the fundamental quantity for the BdGE, the gap.

\begin{figure}[t]
\centering
\includegraphics[width=0.6\textwidth]{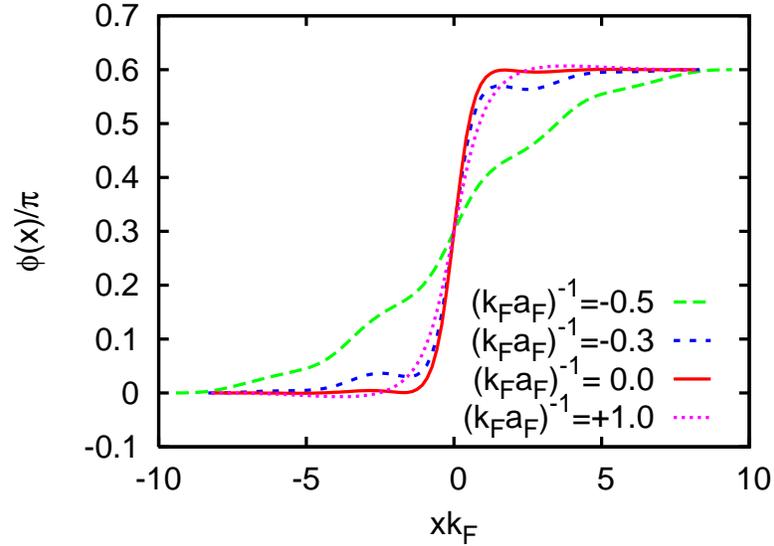}
\caption{Crossover soliton phase $\phi(x)$ corresponding to the gap amplitude shown in Fig.~\ref{fig:1}.\label{fig:4}}
\end{figure}
Figure~\ref{fig:4} illustrates the characteristic phase profile of crossover solitons.  Such a phase profile was not accessible to previous work where only real solutions were considered~\cite{antezza2007}; here we find general complex single-soliton solutions.  For all stationary solutions to the NLSE, including dark solitons, the slope of the phase is proportional to the inverse density~\cite{carr2000a}.  Thus the phase follows the density and strong features in the density show up also in the phase.  For crossover solitons we observe a similar behavior in the phase of the gap. For example, Friedel oscillations are also apparent in the phase, and close examination of Figs.~\ref{fig:1} and~\ref{fig:4} reveals that such oscillations appear for the same values of $x k_F$.  This gives hope in the future to find an analytical relationship between $|\Delta|^2$ and phase for crossover solitons analogous to the existing relationship for NLSE dark solitons.  We are not aware of any method to measure the gap phase; the connection we have found between gap and density modulations in an extended region of the BCS-BEC crossover should, however, allow one to extract this information from density modulations in an interference experiment, similar to what is done in BECs~\cite{andrews1997}.

\begin{figure}[t]
\centering
\includegraphics[width=0.35\textwidth]{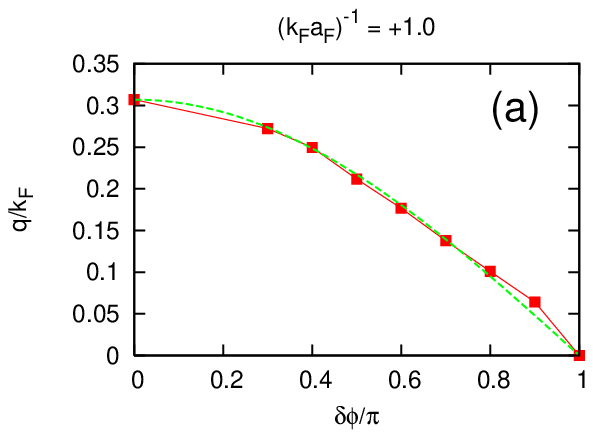}\hspace*{0.2cm}
\includegraphics[width=0.35\textwidth]{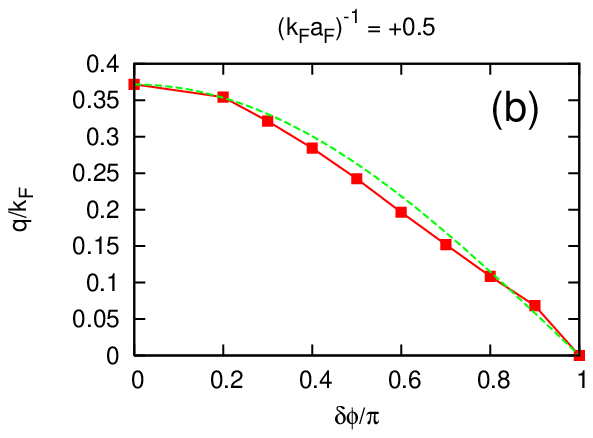}\\
\vspace*{0.2cm}
\includegraphics[width=0.35\textwidth]{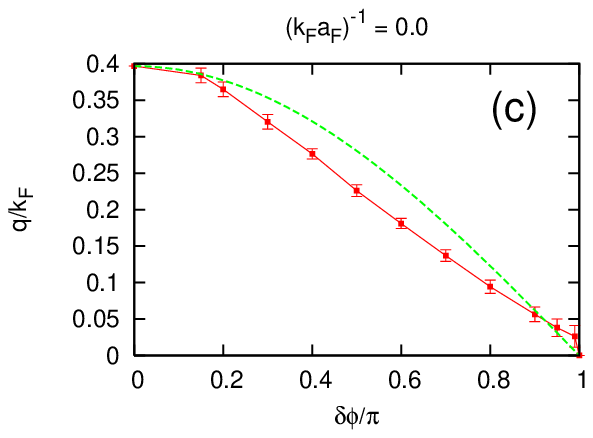}\hspace*{0.2cm}
\includegraphics[width=0.35\textwidth]{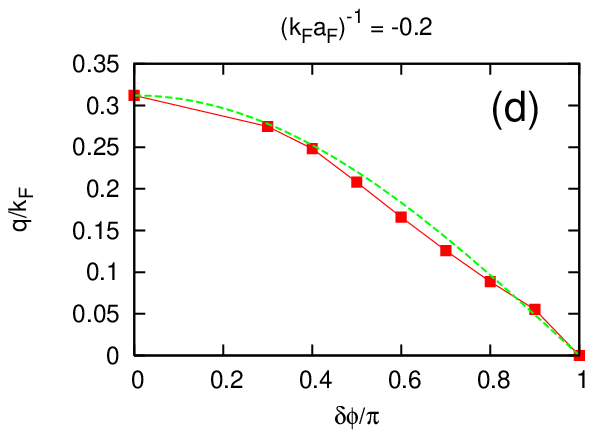}
\caption{Soliton velocity vs.~phase difference
$\delta\phi$ for different coupling values $(k_F a_F)^{-1}$.  The solid red curves are a guide to the eye, while red square points represent actual calculated data.  The green dashed curve shows the NLSE prediction for comparison, with $v_c$ set to the Landau critical velocity for the corresponding value of $(k_F a_F)^{-1}$.\label{fig:5}}
\label{v_vs_phiUL}
\end{figure}
\begin{figure}[t]
\centering
\includegraphics[width=0.6\textwidth]{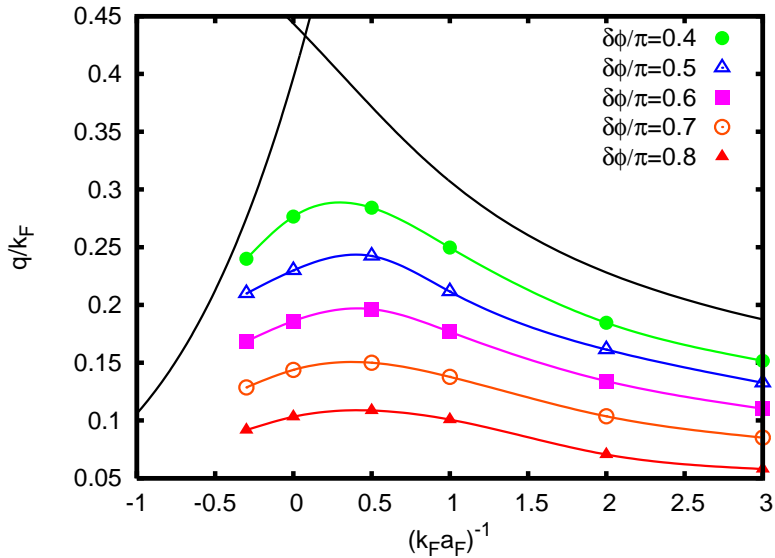}
\caption{Soliton velocity for several values of the phase difference
$\delta\phi$ vs.~coupling $(k_F a_F)^{-1}$; points represent actual data while curves are a guide to the eye. The black curves
correspond to the two branches of the Landau critical velocity, the depairing velocity to the left and the sound mode velocity to the right, and are the
expected limiting curves for $\delta\phi\to 0$.\label{fig:6}}
\end{figure}

Figure~\ref{fig:5} shows the soliton velocity vs.~the phase difference $\delta\phi$ for four characteristic coupling strengths.
These plots clearly show that the curve $v(\delta\phi)$ deviates from the $\cos(\delta\phi/2)$ most strongly at the unitarity limit, where it becomes
almost linear. In the BEC limit an analytical relation is given by Eq.~(\ref{eq:josephson}). In the weak-coupling limit we do not have a similar
analytic solution, but Fig.~\ref{fig:5}(d) suggests that nearly the same functional form is valid in weak coupling.
For all plots, the point corresponding to $\delta\phi=0$ was obtained by the independent calculation of the Landau critical velocity for
a homogeneous superfluid through the crossover discussed in Ref.~\cite{spuntarelli2007}. The point at $\delta\phi=\pi$ is the black soliton case, a purely real solution up to an arbitrary constant phase, which is reproduced from Ref.~\cite{antezza2007}.
The matching of the data at finite $\delta\phi$ with the value at $\delta\phi=0$ provides a check of the calculation.
The curves in Fig.~\ref{fig:5} also show that the maximum allowed velocity for a soliton is given by the Landau critical velocity, which is in turn determined by the lesser of the sound mode and depairing velocities (see also Fig.~\ref{fig:6}). Note in this respect that the depairing velocity would be completely missed in approaches based
on the use of superfluid hydrodynamic equations~\cite{wenW2009}.

Our results in Fig.~\ref{fig:5} are accurate to approximately a few percent, and the error bars are approximately given by the size of the data points as depicted.  Close to $\delta\phi=\pi$, a crossing occurs between the NLSE prediction (green dashed curve) and the data (red solid curve), which seems to originate from a small reentrant behavior of the velocity vs.~$\delta\phi$ curve in this region.  The reentrant behavior is most pronounced at unitarity, where the deviation from the NLSE prediction is also strongest.  For this case, illustrated in Fig.~\ref{fig:5}(c), we have decreased the point size and presented a careful study of the error bars; see Sec.~\ref{sec:method} for a description of how we determine the error.  Unfortunately, the error of our numerical method is also largest in the reentrant region as $\delta\phi/\pi \to 1^-$.  Our error bars indicate that reentrant behavior is a real physical effect at least to the one sigma level.  This point bears further investigation in the future.

Finally, in Fig.~\ref{fig:6} we present a study of crossover soliton velocity for
both different interaction strengths and phase differences. It is clearly apparent that as $\delta\phi\to 0$ the velocity approaches the lesser of the depairing velocity and the sound mode velocity.  Thus all crossover soliton characteristics are strongly constrained by these basic and well understood velocity scales in the problem, as apparent in previous figures.

\section{Discussions and Conclusions}
\label{sec:conclusions}

We have explored crossover solitons from weak to strong interactions via mean field theory, i.e., the Boguliubov-de Gennes equations.  In an interacting Fermi gas, such solitons are complex solutions of the BdGE appearing as a notch in the gap amplitude with a characteristic phase difference.  We find that the phase follows closely the gap amplitude, for instance in the location and strength of Friedel oscillations.  The gap is not presently observable in strongly interacting Fermi gases, but the density is.  We found that crossover solitons are evident in the density as well as in the gap.  In the strongly interacting limit Fermi pairs are tightly bound, the system is effectively a weakly interacting BEC from the bosonic perspective, and an effective NLSE description holds.  Single dark soliton solutions to the NLSE have a characteristic Josephson-like sinusoidal relationship between velocity and phase difference across the notch.  In contrast, we find that crossover solitons have a near-linear relationship between velocity and phase difference at unitarity.  We find the relation between the gap and the density is close to a simple power law, which is quadratic in the BEC limit and drops off strongly as the interaction strength is decreased; at unitarity the exponent is about 1.5.  In general, all crossover soliton characteristics are strongly constrained by the depairing velocity for weak interactions on the BCS side and the sound mode velocity for strong interactions on the BEC side. This feature is completely missed in approaches based on the use of superfluid hydrodynamic equations~\cite{wenW2009}, which are sensitive only to the sound mode velocity and therefore break down on the BCS side of the crossover. Only the full self-consistent solution of the Boguliubov-de Gennes equations is able to
take into account both Friedel oscillations and depairing effects in crossover solitons. The solution of this numerically demanding problem was possible by the application of the same methodology and numerical strategies, developed by three of us~\cite{spuntarelli2007,spuntarelli2010} for the study of the Josephson effect, to the crossover soliton problem. The soliton problem and the Josephson effect in the BCS-BEC crossover have been placed, in this way, within the same theoretical framework, both conceptually and practically.

Crossover solitons can be made in experiments on ultracold fermions in a similar way to how vortices are sometimes made in such experiments.  A dark soliton can be created on the BEC side of the crossover, and then the Fano-Feshbach resonance tuned towards unitarity or the BCS side.  Although vortices are made by imposing rotation on a BEC, dark solitons are made via phase and density engineering, as described in some detail in previous work by one of us~\cite{carr2001e}.  The basic concept is to create a density notch via a tightly focused Gaussian beam.  The intensity of the beam is increased adiabatically until the desired depth of the notch is achieved.  Then the beam is turned off suddenly.  At the same time as the beam is turned off a phase imprint is imposed with another beam, acting as an impulse.  Various combinations of these techniques have been used in previous BEC dark soliton experiments discussed in Sec.~\ref{sec:introduction}.  Both experimental and theoretical aspects of creation of band or planar solitons, as we have treated in this Article, is discussed by one of us in Ref.~\cite{carr2008f,carr2008g}.

An important open question remains about dark solitons through the crossover.  Are they in fact solitary waves?  At least deep on the BEC side, a number of studies taking into account quantum fluctuations (e.g.,~\cite{martinAD2010}) indicate that, although dark solitons undergo quantum delocalization, they continue to collide elastically.  On the other hand, in a lattice with small filling factors of 1 to 3 atoms per site there is evidence that solitons become entangled and decay, even without collisions~\cite{carr2009l,carr2009m}.  Since a lattice effectively induces quantum fluctuations, it may be that this is indicative of quantum instabilities not properly represented by the NLSE limit of the BdG equations.  It is an intriguing question  whether or not the BdG equations can describe such fluctuations for $(k_F a_F)^{-1}\lesssim 1$, and if solitons collide elastically in these circumstances.  Within mean-field theory, we expect that our solutions should have essentially the same decay channels as the corresponding solitonic solutions of the NLSE, i.e., snake instability and creation of vortices. These instabilities do not prevent the creation and observation of sufficiently long-lived solitons in a BEC~\cite{burger1999,denschlag2000,anderson2001}.  We expect the same to hold for a unitary superfluid Fermi gas.

Another interesting open problem is the likelihood of a metastable finite-size quantum phase transition for a crossover soliton on a ring.  For fixed interaction strength and number of fermions, a sufficiently small ring will not have a stationary crossover soliton solution.  As the interaction strength, number of fermions, ring size, and/or rotation induced in the system are varied, such a solution should become possible, leading to a finite-size phase transition as explored in a number of papers for the Lieb-Liniger Hamiltonian for interacting bosons~\cite{carr2008a,carr2009f,carr2010c,carr2010h}.  As there is an exact mapping from bosons to fermions, one expects the same kind of transition to occur in the fermionic system in 1D.  An interesting question is whether this phase transition appears also in 3D under our assumption of spatial uniformity in the transverse directions.

Finally we briefly mention that if there are unpaired fermions, the crossover soliton problem is much more general, with a possible Fulde-Ferrel-Larkin-Ovchinikov (FFLO) phase~\cite{liaoYA2010}.  In certain regimes one can anticipate that unpaired fermions would prefer to occupy the density-notch region of the crossover soliton, and a crossover analogy to the bright-in-dark BEC soliton~\cite{middelkamp2010} will become possible.

We acknowledge discussions with Joachim Brand.  L.D.C. and P.P. acknowledge support from the Kavli Institute for Theoretical Physics in Santa Barbara and the Institut Henri Poincar\'e in Paris.  L.D.C. acknowledges support from the National Science Foundation under Grant PHY-0547845 as part of the NSF CAREER program. P.P. and G.C.S. acknowledge support from the Italian MIUR under Contract Cofin-2007 ``Ultracold Atoms and Novel Quantum Phases.''

After submission of this work, two related studies appear by our colleagues, which can be read along with this Article to present different perspectives on the very interesting question of the physical properties and characterization of crossover solitons~\cite{scottRG2010,liaoR2010}.

\end{document}